\newcommand{\titext}{Mass-Varying Massive Gravity with k-essence}
\newcommand{\abstext}{
For a large class of mass-varying massive gravity models, the graviton mass cannot provide the late-time cosmic expansion of the universe due to its vanishing at late time.  In this work, we propose a new class of mass-varying massive gravity in which the graviton mass varies according to a kinetic term of a k-essence field. By using a more general form of the fiducial metric, we found a solution such that a non-vanishing graviton mass can drive the accelerated expansion of the universe at late time. We also perform dynamical analyses of such model and found that without introducing the k-essence Lagrangian, the graviton mass can be responsible for both dark contents of the universe, namely dark energy that drives the accelerated expansion of the universe and non-relativistic matter that plays the role of dark matter. Moreover, by including the k-essence Lagrangian, we found that it is possible to alleviate the so-called cosmic coincidence problem.
}

\documentclass[10pt,english,preprint]{revtex4-1}

\usepackage[T1]{fontenc}
\usepackage{amsmath}
\usepackage{amssymb}
\usepackage{esint}
\usepackage{graphicx}
\usepackage{caption}
\usepackage{subcaption}
\usepackage{bm}
\usepackage{color}



\begin{document}

\title{\titext}
\author{Lunchakorn Tannukij}

\affiliation{Department of Physics, Faculty of Science,  Mahidol University, Bangkok 10400, Thailand}
\author{Pitayuth Wongjun}
\affiliation{The Institute for Fundamental Study, Naresuan University,
Phitsanulok 65000, Thailand}

\affiliation{Thailand Center of Excellence in Physics, Ministry of Education,
Bangkok 10400, Thailand}

\begin{abstract}
\abstext
\end{abstract}

\maketitle

\flushbottom

\section{Introduction}

Massive gravity has its own series of developments as a modified gravity beyond general relativity. Back in 1939, Fierz and Pauli investigated a first model of massive gravity \cite{Fierz:1939ix}. The model was a linearized general relativity, where the fluctuation of geometry propagates a spin-2 graviton, plused linear interactions which, in a particle physics language, corresponds to giving a non-zero mass to the graviton, hence the name ``massive gravity''.  This model was supposed to be coincide with general relativity in a massless limit but it faced a theoretical crisis when  discontinuities in such limit were found by van Dam, Veltman, and Zakharov \cite{vanDam:1970ab,Zakharov:1970cd}. In particular, the discontinuities were found as different predictions between Fierz-Pauli massive gravity and general relativity. The problem remained for several years until Vainshtein proposed a way out by introducing higher-order interactions into the Fierz-Pauli massive gravity \cite{Vainshtein:1972sx}. In other words, he claimed that inside a particular scale, coined Vainshtein radius, any predictions from the linear theory cannot be trusted unless nonlinear contributions are taken into account. However, adding such nonlinearities, claimed by Boulware and Deser, not only fixes the discontinuity problem but also introduces a theoretical inconsistency, namely a Boulware-Deser ghost \cite{Boulware:1973my}. This ghost is an extra degree of freedom, apart from 5 degrees of freedom  originally existing in the linear massive gravity, whose kinetic term has a wrong sign. The ghost problem had been a blockage for the massive gravity theory until recently in 2010, de Rham, Gabadadze, and Tolley found suitable nonlinear interactions which does not excite the Boulware-Deser mode, dubbed dRGT massive gravity \cite{deRham:2010ik,deRham:2010kj}. Thus, massive gravity became again an active field of study.

Although it was just a generalization back then, massive gravity has its modern motivations. Introducing a non-zero mass to a graviton shrinks the scale at which the gravity works. In other words, the graviton mass weakens the gravitation at large scale.  As a result, it allows a cosmic acceleration and hence may be able to describe the mysterious dark energy in its language. This motivates cosmologists to study its cosmological implications. Moreover, since de Rham, Gabadadze, and Tolley found the healthy nonlinear massive gravity, the theory had again opened a door to various researches on massive gravity, not only its cosmology but people also study astrophysical objects in the theory like black holes \cite{Volkov:2013roa,Tasinato:2013rza,Babichev:2015xha,Ghosh:2015cva,Tolley:2015ywa,Ayon-Beato:2015qtt}. For cosmological models of massive gravity, it has been found that the solutions in the models with Minkowski fiducial metric do not admit the flat and closed FLRW solutions for the physical metric \cite{D'Amico:2011jj,Gumrukcuoglu:2011ew}. In order to obtain the all kinds of FLRW solutions, one may consider a general form of the fiducial metric \cite{Fasiello:2012rw,Langlois:2012hk,Gumrukcuoglu:2011zh,Langlois:2013cya,Chullaphan:2015ija}.

It has been found, however, that there are some inconsistencies when cosmology is taken into account. For example, some degrees of freedom cease to exist when the Friedmann-Lema\^itre-Robertson-Walker (FLRW) ansatz is assumed \cite{Gumrukcuoglu:2011zh}. This leads to numerous studies beyond the dRGT massive gravity \cite{DeFelice:2012ac,Mukohyama:2012op,Antonio:2013qr,Huang:2012pe,Wu:2013ii,Leon:2013qh,Huang:2013mha,D'Amico:2012zv,Gumrukcuoglu:2013nza,D'Amico:2013kya,DeFelice:2013za,DeFelice:2013gm,Heisenberg:2015voa,Kahniashvili:2014wua,Gumrukcuoglu:2014xba,Solomon:2014iwa,Hinterbichler:2013dv,Gabadadze:2012tr,Andrews:2013uca}. One of those is to generalize a constant graviton mass to be varied by other scalar field, dubbed mass-varying massive gravity \cite{Huang:2012pe,Wu:2013ii,Leon:2013qh,Huang:2013mha}. The theory is  proven to be free of Boulware-Deser ghost. However, cosmological implications of such model indicates the universe with subdominant contributions from massive gravity. In particular, the graviton mass is governed by an inverse of a scale factor of the universe which will vanish at late time. Consequently, such model cannot give a proper explanation to the cosmic expansion caused by the massive graviton.

In this work, we propose an alternative way to construct a mass-varying massive gravity. Not only by a scalar field, but the graviton mass is also determined by the kinetic term of the scalar field. Moreover, the scalar field is governed by a k-essence Lagrangian \cite{ArmendarizPicon:2000dh, ArmendarizPicon:2000ah,Chiba:1999ka}. Under the FLRW ansatz, we found a solution whose the graviton mass do not necessarily vanish at late time. Moreover, by assuming both the k-essence and the graviton mass behave as  perfect fluids, we found that the graviton mass can give rise to  a ``dust-like'' matter while combining with other contributions it is possible to have an equation of state parameter close to $-1$, as suggested by  recent observation \cite{Ade:2013zuv}. Such matter may be responsible for a dark matter, another mysterious content known to exist in addition to the ordinary matter. Since the graviton mass can give rise to both of the dark contents, it is tempting to consider its evolution whether there exists an epoch in which both contents in the dark sector are comparable, the so-called cosmic coincidence problem.

Our paper is organized as follows. In  section \ref{sec:eom}, the proposed model is addressed along with its  equations of motion in the FLRW background. We also discuss some crucial properties of the model in this section where we have shown the existence of the dust-like matter expected to be responsible for the dark matter.
With helps from appropriate assumptions, we show in section \ref{sec:trial} the solution to this model which corresponds to the dark energy and the non-vanishing characteristic of the graviton mass existing in this model.
 After getting some perspectives, we begin the dynamical system analyses in section \ref{sec:dyna} to find all possible fixed points and their stabilities and the extended analyses are covered in section \ref{sec:exdyna}. We conclude our work in the last section by the discussion on key ideas of our work and on whether or not the coincidence  problem is alleviated.

\section{The model and the background equations}\label{sec:eom}
We consider a mass-varying dRGT massive gravity action where the graviton mass is varied by the k-essence field. Usually, one may consider a graviton mass as a function which varies as the scalar field propagates \cite{Huang:2012pe,Wu:2013ii,Leon:2013qh,Huang:2013mha}. However, in this work, we will consider the graviton mass not only as a function of the scalar field $\phi$ but also its kinetic term $X \equiv -\frac{1}{2}g^{\mu\nu}\nabla_\mu \phi \nabla_\nu \phi$.  The action of such model can be expressed as
\begin{align}
S &= \int d^4 x \sqrt{-g} \bigg[ \frac{M_p^2}{2} R[g] + V(X,\phi) (\mathcal{L}_2[g,f] +\alpha_3\mathcal{L}_3[g,f] +\alpha_4\mathcal{L}_4[g,f] )  \nonumber
\\
&\qquad\qquad\qquad+ P(X,\phi)\bigg]. \label{eqn:action}
\end{align}
where $R$ is a Ricci scalar corresponding to a physical metric $g_{\mu\nu}$, $V(X,\phi)$ is a square of the graviton mass which depends on the scalar field and its kinetic term, $\mathcal{L}_i$'s represent the interactions of the $i$th order of the massive graviton, and $P(X,\phi)$ is a Lagrangian of the k-essence field. In particular, those interactions of the massive graviton are constructed from two kinds of metrics and can be expressed as follows,
\begin{align}
\mathcal{L}_2[g,f]&=\frac{1}{2}\left([\mathcal{K}]^2-[\mathcal{K}^2]\right), \label{L2}
\\
\mathcal{L}_3[g,f]&=\frac{1}{3!}\left([\mathcal{K}]^3-3[\mathcal{K}][\mathcal{K}^2]+2[\mathcal{K}^3]\right), \label{L3}
\\
\mathcal{L}_4[g,f]&=\frac{1}{4!}\left([\mathcal{K}]^4-6[\mathcal{K}]^2[\mathcal{K}^2]+3[\mathcal{K}^2]^2+8[\mathcal{K}][\mathcal{K}^3]-6[\mathcal{K}^4]\right), \label{L4}
\end{align}
where the tensor $\mathcal{K}_{\mu\nu}$ is constructed from the physical metric $g_{\mu\nu}$ and an another metric $f_{\mu\nu}$ as
\begin{align}
\mathcal{K}^{\mu}_{\;\nu} = \delta^\mu_{\;\nu} - \left(\sqrt{g^{-1}f}\right)^{\mu}_{\;\nu}.
\end{align}
where the square roots of those tensors are defined so that $\sqrt{g^{-1}f}^{\mu}_{\;\rho} \sqrt{g^{-1}f}^{\rho}_{\;\nu} = \left(g^{-1}f\right)^{\mu}_{\;\nu}$.
In massive gravity, apart from the physical metric, there exists another kind of the metric tensor $f_{\mu\nu}$, usually named ``fiducial metric'', which is an object introduced to the theory so that one can construct non-trivial interactions from metric tensors as in Eq. (\ref{L2}), (\ref{L3}), and (\ref{L4}). Those complicated combinations in the interactions, with arbitrary values of the parameters $\alpha_3,\alpha_4$, are to ensure the absence of the Boulware-Deser (ghostly) degree of freedom \cite{deRham:2010ik,deRham:2010kj}. Moreover, thanks to the Stuckelberg tricks, the general covariance, or the gauge symmetry, can be well integrated  into the massive gravity via
\begin{align}
f_{\mu\nu} = \partial_\mu \varphi^\rho \partial_\nu \varphi^\sigma \tilde{f}_{\rho\sigma},
\end{align}
provided that each of the fields $\varphi^\mu$'s transforms as scalar under any coordinate transformation. As for the $\tilde{f}_{ab}$, one can choose it to be any kind of metric which shares the same symmetries as the physical metric does. For example, one can have a 4-dimensional Minkowski metric to be the fiducial metric for a cosmological solution \cite{Gumrukcuoglu:2011ew}, or even a higher-dimensional kind of metric whose the reduced 4-dimensional metric is isotropic and homogeneous is considered as the fiducial metric in the cosmological solution \cite{Chullaphan:2015ija}.

In this work, we consider the cosmological implications of the proposed model, where the isotropic and homogeneous universe is assumed  whose spacetime is represented quite well by the Friedmann-Lema\^itre-Robertson-Walker (FLRW) metric  as follows,
\begin{align}
ds^2 = -N(t)^2 dt^2 + a(t)^2 \Omega_{ij}(x) dx^i dx^j,
\end{align}
where $N(t)$ is a lapse function, $a(t)$ represents a scale factor which determines the scale of the spatial distance, and
\begin{align}
\Omega_{ij}(\varphi) = \delta_{ij} + \frac{k\delta_{ia}\delta_{jb}\varphi^a \varphi^b}{1-k \delta_{lm}\varphi^l \varphi^m},
\end{align}
is the spatial maximally symmetric metric whose the spatial curvature is characterized by $k \in \{-1,0,+1\}$ corresponding to the open, flat, and closed geometry respectively. As claimed, the FLRW ansatz is also used as the fiducial metric,
\begin{align}
\tilde{f}_{\mu\nu}d\varphi^\mu d\varphi^\nu &= -n(\varphi^0)^2\left(d\varphi^0\right)^2 + \alpha(\varphi^0)^2 \Omega_{ij}(\varphi) d\varphi^i d\varphi^j,
\\
f_{\mu\nu} &=\partial_\mu \varphi^\rho \partial_\nu \varphi^\sigma \tilde{f}_{\rho\sigma},
\end{align}
where $n$ and $\alpha$ are a lapse function and a scale factor in the fiducial sector.
Plugging those in  Eq. (\ref{eqn:action}), the mini-superspace action of the model reads
\begin{align}
S &= \int d^4 x \sqrt{\frac{1}{1-kr^2}} \bigg[ M_p^2 \left(-3\frac{a\dot{a}^2}{N}+3kNa\right) + 3Na^3V\left(F-G\frac{n}{N}\right) +Na^3P \bigg], \label{eqn:FLRWaction}
\end{align}
where
\begin{align}
F &\equiv \left(2+\frac{4}{3}\alpha_3+\frac{1}{3}\alpha_4\right) -\left(3+3\alpha_3+\alpha_4\right)\bar{X} +\left(1+2\alpha_3+\alpha_4\right)\bar{X}^2 -\left(\alpha_3+\alpha_4\right)\frac{\bar{X}^3}{3},
\\
G &\equiv \frac{1}{3}\left(3+3\alpha_3+\alpha_4\right) -\left(1+2\alpha_3+\alpha_4\right)\bar{X} +\left(\alpha_3+\alpha_4\right)\bar{X}^2 -\alpha_4\frac{\bar{X}^3}{3},
\end{align}
and we have defined
\begin{align}
\bar{X} \equiv \frac{\alpha}{a}, \quad \eta \equiv \frac{n}{N}.
\end{align}
To determine the dynamics of the system, one can vary the action in Eq. (\ref{eqn:FLRWaction}) with respect to dynamical variables which are $N$, $a$, $\phi$, and the Stuckelberg fields $\varphi^\mu$. The corresponding equations of motion, assuming the unitary gauge $\varphi^\mu = x^\mu$ for simplicity, read
\begin{align}
M^2_p \left(3H^2+3\frac{k}{a^2}\right) &= -3VF + 6XV_{,X} \left(F-G\eta\right) + \left(2XP_{,X}-P\right), \label{eqn:neom}
\\
M^2_p\left(\frac{2\dot{H}}{N}+3H^2+\frac{k}{a^2}\right) &= -3VF+VF_{,\bar{X}}\left(\bar{X}-\eta\right)-P, \label{eqn:aeom}
\\
\frac{\dot{V}}{V} &= NH\left(1-h \bar{X}\right)\frac{F_{,\bar{X}}}{G}, \label{eqn:stuckeom}
\\
Na^3\left(3V_{,\phi}\left(F-G\eta\right)+P_{,\phi} \right) &= \frac{d}{dt}\left[\left(a^3\sqrt{2X}\right)\left(3V_{,X}\left(F-G\eta\right)+P_{,X}\right)\right], \label{eqn:conserve1}
\\
3HN(-2XP_{,X}-6XV_{,X}\left(F-G\eta\right) &+ VF_{,\bar{X}}\left(\bar{X}-\eta\right))\nonumber
\\
&= \frac{d}{dt} \left(-3VF + \left(2XP_{,X}+6XV_{,X}\left(F-G\eta\right)\right)-P\right) \label{eqn:conserve2},
\end{align}
where the last equation is obtained from the conservation on the energy-momentum tensor; $\nabla_{\mu} T^\mu_{\;\nu}=0$ and we have defined
\begin{align}
h \equiv \frac{H_\alpha}{H}, \quad H_\alpha \equiv \frac{\dot{\alpha}}{\alpha n}.
\end{align}
From the above equations, one can see that Eq. (\ref{eqn:neom}) is a Friedmann equation with extra matter contents coming from the graviton mass. As a partner to the Friedmann equation, the so-called acceleration equation corresponds to Eq. (\ref{eqn:aeom}). Since we have the Bianchi's identity relating the equations of motion, these 5 equations of motion are not entirely independent.
Note that this set of equations recovers the original self-accelerating cosmology when the square of a graviton mass $V$ is constant by which the usual condition $F_{,\bar{X}}\left(1-h\eta\right)$ is obtained readily from Eq. (\ref{eqn:stuckeom}) \cite{Gumrukcuoglu:2011ew}.
However, as $V$ being constant is no longer the case, the equations of motion look even more complex than those in general relativity. To simplify the following calculations, we choose $P$ such that the k-essence field behaves as perfect fluid. The appropriate form of $P$ which satisfy such behaviour is
\begin{align}
P(X,\phi) = P_0 X^{\frac{1+w}{2w}} = P_0 X^{\gamma/2}, \label{Pperfect}
\end{align}
where $\gamma \equiv 2XP_{,X}/P \equiv \frac{1+w}{2w}$, $P_0$ is a constant, and $w$ is an equation of state parameter corresponding to the perfect fluid represented by the k-essence field \cite{Boubekeur:2008kn}. Moreover, we let the graviton mass function mimics the perfect-fluid form as
\begin{align}
V = V_0  X^{\lambda/2}, \label{Vperfect}
\end{align}
whose $\lambda$ characterizes the power of the kinetic term as $\gamma$ does for $P$, i.e. $\lambda \equiv 2XV_{,X}/V$ and $V_0$ is a constant. Note that under these assumptions, both $P$ and $V$ vary according to the kinetic term of $\phi$ but not the $\phi$ itself.  Usually, in the quintessence model the continuity equation for the scalar field is obtained from the equation of motion of $\phi$ \cite{Ratra:1987rm,Wetterich:1987fm}. Taking that into account, we consider the equation of motion of $\phi$ in Eq. (\ref{eqn:conserve1}), then under the perfect-fluid assumptions for $P$ and $V$ in Eq. (\ref{Pperfect}) and Eq. (\ref{Vperfect}) we have
\begin{align}
\frac{d}{dt}\left(\left(\frac{a^3}{\sqrt{2X}}\right) \left(6XV_{,X}\left(F-G\eta\right)+2XP_{,X}\right)\right) &= 0. \label{eqn:scalar}
\end{align}
After simple manipulations, the above equation gives the continuity equation for the k-essence field as
\begin{align}
\frac{d}{d t}\rho_X + 3HN  \rho_X
= \frac{\dot{X}}{2X}\rho_X, \label{darkmatter}
\end{align}
where we have defined
\begin{align}
\rho_X \equiv \left(2XP_{,X}+6XV_{,X}\left(F-G\eta\right)\right).
\end{align}
Eq. (\ref{darkmatter}) determines the dynamics of the matter of energy density $\rho_X$ which resides in the Friedmann equation in Eq. (\ref{eqn:neom}).
Interestingly, this looks exactly like a continuity equation of a ``dust-like'' matter  with an interaction to the other matter sector determined by the flow rate of the form $\frac{\dot{X}}{2X}\rho_X$.  One can also integrate Eq. (\ref{eqn:scalar}) to find an expression for $\rho_X$ in terms of the scale factor as
\begin{align}
\rho_X = \frac{\sqrt{2X}C}{a^3}.
\end{align}
where $C$ is an integration constant. In case of a constant $X$, this ensures one of the properties that this matter shares with the dust; the energy density is as inversely proportional to  $a^3$ as the dust is. According to such characteristics, it is reasonable to interpret $\rho_X$ as a dark matter. By doing so, this kind of dark matter possesses some interesting features. Firstly, $\rho_X$ is a dust-like matter which  can arise naturally from the massive gravity sector indicating that dark matter may be just an artifact of the varying graviton mass caused by the kinetic term of the k-essence field. Moreover, this claim is still valid even in the case of $P=0$.
Since a graviton mass can represent dark energy in a generic class of the dRGT massive gravity, this suggests a unification of the dark sector, namely dark energy and dark matter, by such varying graviton mass.
Secondly, by having this kind of matter in the theory, we may expect this model of mass-varying massive gravity to solve the cosmic coincidence problem, where the universe is known to be composed mainly of comparable amounts of dark energy and dark matter. Thanks to the unification suggested above, it may be possible to provide an explanation on the coincidence problem by the existence of the graviton mass alone while the cosmic acceleration also counts.

Since the equations of motion are coupled in a very cumbersome way, to get a whole picture of this system we need to perform a dynamical analysis, which is the main subject in the very last section.  However, we can still get some rough descriptions, as a guideline to the dynamical analysis, by introducing some simple assumptions to the system, which is done in the next section.

\section{Dark energy solution for the self-accelerating universe}\label{sec:trial}
It is widely known that our universe is expanding with acceleration for which dark energy is responsible. There is recently an observational evidence  indicating that the observed effective equation of state parameter of the dark energy is close to $-1$ \cite{Ade:2013zuv}. In this section, we shall adopt this characteristic by treating all the contributions from the graviton mass to have such property. We define
\begin{align}
\rho_g &\equiv -3VF +6XV_{,X}\left(F-G\eta\right), \label{eqn:rhog}
\\
p_g &\equiv 3VF-VF_{,\bar{X}}\left(\bar{X}-\eta\right). \label{eqn:pg}
\end{align}
From the above definition, the corresponding equation of state parameter is defined as
\begin{align}
w_g\equiv \frac{p_g}{\rho_g}.
\end{align}

By treating $\rho_g$ as an energy density of dark energy, we set $w_g=-1$ and then we have the following condition,
\begin{align}
6XV_{,X}\left(F-G\eta\right) &= VF_{,\bar{X}}\left(\bar{X}-\eta\right). \label{matchingeos}
\end{align}
To simplify the calculation, we use the perfect-fluid form of $V$ in Eq. (\ref{Vperfect}). Consequently, Eq. (\ref{matchingeos}) becomes
\begin{align}
3\lambda\left(F-G\eta\right) &= F_{,\bar{X}}\left(\bar{X}-\eta\right),
\\
\lambda &= \frac{F_{,\bar{X}}\left(\bar{X}-\eta\right)}{3\left(F-G\eta\right)}. \label{sol:exponent}
\end{align}
Eq. (\ref{sol:exponent}) is a requirement for the exponent $\lambda$ to have a solution with the equation of state equal to $-1$. To get a picture of this characteristic, let us assume
\begin{align}
\bar{X} &= \text{constant},
\\
\eta &= \text{constant},
\\
\text{then}\quad h &= \frac{1}{\eta}.
\end{align}
Under these assumptions, the exponent $\lambda$ in Eq. (\ref{sol:exponent}) is just a constant. To investigate this further, we consider Eq.  (\ref{eqn:stuckeom}) under the previous assumptions,
\begin{align}
\frac{\dot{V}}{V} &=NH\left(1-h \bar{X}\right) \frac{F_{,\bar{X}}}{G},\nonumber
\\
\frac{\lambda\dot{X}}{2X}&= NH\left(1-\frac{\bar{X}}{\eta} \right)\frac{F_{,\bar{X}}}{G},\nonumber
\\
&= -\left(\bar{X}-\eta \right)\frac{F_{,\bar{X}}}{G\eta} \frac{\dot{a}}{a}.
\end{align}
From the condition of $\lambda$ in Eq. (\ref{sol:exponent}),
\begin{align}
\frac{\dot{X}}{X} = -\frac{6\left(F-G\eta\right)}{G\eta} \frac{\dot{a}}{a}.
\end{align}
Since $\bar{X}, \eta$, and hence $F$ and $G$ are constant, this equation can be integrated easily,
\begin{align}
\int \frac{dX}{X} &= -\frac{6\left(F-G\eta\right)}{G\eta} \int \frac{da}{a},\nonumber
\\
X &= C_0 a^{-\frac{6\left(F-G\eta\right)}{G\eta}} \label{sol:X}
\end{align}
where $C_0$ is an integration constant.
Now we have
\begin{align}
V = V_0 X^{-\frac{\left(1-\frac{\bar{X}}{\eta}\right)\eta F_{,\bar{X}}}{6\left(F-G\eta\right)}} = V_0 C_0 a^{\left(1-\frac{\bar{X}}{\eta}\right)\frac{F_{,\bar{X}}}{G}}. \label{eqn:dedmass}
\end{align}
Furthermore, Eq. (\ref{sol:X}) can possibly determine a relation between the scale factor and the rate of change of the scalar field since
\begin{align}
X = \frac{\dot{\phi}^2}{2N^2} = C_0 a^{-\frac{6\left(F-G\eta\right)}{G\eta}}. \label{eqn:kineticsol}
\end{align}

The expression of $V$ in Eq. (\ref{eqn:dedmass}) shows the evolution of the (square of) graviton mass as $a$ evolves. In the previous model of mass-varying massive gravity \cite{Huang:2012pe,Wu:2013ii,Leon:2013qh,Huang:2013mha}, in which the Minkowski fiducial metric is used, the varying graviton mass shrinks as the scale factor grows. In this model, however, the exponent in Eq. (\ref{eqn:dedmass}) determines  whether the graviton mass will shrink or not as the scale factor grows, or remain constant in case that the exponent vanishes. Note that this crucial difference is caused by the different form of the fiducial metric, which is the FLRW metric in this case compared with the Minkowski one in the previous models. This result indicates the sensitivity of the fiducial metric existing in the generic dRGT massive gravity where different fiducial metrics set different stages for the system and provide different solutions \cite{Fasiello:2012rw,Langlois:2012hk,Gumrukcuoglu:2011zh,Langlois:2013cya,Chullaphan:2015ija}.

One more crucial point of this analysis is that the contributions from the graviton mass can have the same equation of state parameter as dark energy while one of those contributions possesses characteristics of dust, namely the term $6XV_{,X}\left(F-G\eta\right)$. From Eq. (\ref{darkmatter}), such term belongs to the dark matter $\rho_X$. This may be a way out for the cosmic coincidence problem since we may infer that varying graviton mass is responsible for a dark matter via the term like $6XV_{,X}\left(F-G\eta\right)$, as we have claimed in the previous section, while it can still drive the accelerating expansion. To verify this idea, and to seek for a finer description of this model, we will perform a dynamical analysis, which is in the next section.

\section{Dynamical system}\label{sec:dyna}
In this section, we will consider dynamics of the universe governed by this new class of mass-varying massive gravity using the method of the autonomous system. Due to the complexity of the graviton mass, we will begin this section with a simple analysis by considering the flat FLRW where $k=0$ and assuming that $\bar{X},\eta$ are constant over time, thus $h=1/\eta$. From this assumption, the evolution of $X$ is simply determined by Eq. (\ref{eqn:stuckeom}) such that
\begin{align}
X' &= \frac{\dot{X}}{HNX} = \frac{2}{\lambda}\frac{F_{,\bar{X}}}{G}\left(1- h \bar{X} \right) = -\frac{6 s}{ \lambda\, r},\\
\lambda &\equiv \frac{2 X V_{,X}}{V}, \label{Def-lambda}
\end{align}
where the prime denotes the derivative with respect to $\ln a$.  The parameters $r$ and $s$ are constant and defined as
\begin{align}
r \equiv \frac{G \eta }{F}, \quad s \equiv \frac{ F_{,\bar{X}} (\bar{X} - \eta)}{3F}.\label{Def-r-s}
\end{align}
In order to obtain a suitable autonomous system, let us define  dimensionless variables as follows,
\begin{align}
x &= -\frac{F V}{M^2_p H^2},\label{Def-x}\\
z & = -\frac{P}{3 M^2_p H^2}, \label{Def-z}\\
y &= \frac{2 XP_{,X} + 6XV_{,X}F(1-r)}{3 M^2_p H^2} = \frac{\rho_X}{3M^2_p H^2}, \label{Def-y}\\
\gamma &\equiv \frac{2 X P_{,X}}{P}.\label{Def-gamma}
\end{align}
By using these variables, the equations of motion can be written
in the form of autonomous equations as
\begin{align}
x' &= 3 x \left(y+s x -\frac{s}{r}\right),\label{Eq-x}\\
y' &= 3 y\left(y+s x -1-\frac{s}{\lambda r}\right),\label{Eq-y}\\
\lambda' &= \frac{6 s}{r}\left(\frac{\lambda}{2} -(1+\Gamma)\right),\label{Eq-lambda}\\
1 &= x+ y + z, \label{Eq-con1}\\
y &=  -\lambda x (1-r) - z \gamma.\label{Eq-con2}
\end{align}
where $\Gamma \equiv X V_{,XX}/V_{,X}$. Since we have five variables with two constraints, it is sufficient to consider only three equations. Note that the constraint in Eq. (\ref{Eq-con1}) is derived from Eq. (\ref{eqn:neom}) while the constraint in Eq. (\ref{Eq-con2}) is obtained from the definition of $y$ in Eq. (\ref{Def-y}). The equation of $\lambda$ in Eq. (\ref{Eq-lambda}) is not directly depend on the other variables. Therefore, in principle, we can  solve it separately. For simplicity, we can consider $\lambda$ as a parameter and then consider only the autonomous equations with two variables, $x$ and $y$. We will extend our analysis to a more general case with $\lambda$ being the variable in the next section. The effective equation of state parameter can be written in terms of the dimensionless variables as
\begin{eqnarray}
w_{eff} = \frac{P + 3VF - V F_{,\bar{X}}(\bar{X} - \eta)}{3 M^2_p H^2} = -z - x + x s= -1 + y +x s.
\end{eqnarray}
From these autonomous equations, corresponding fixed points can be found by evaluating  $x' =0$ and  $y'=0$ in Eq. (\ref{Eq-x}) and Eq. (\ref{Eq-y}) respectively. Properties of all the fixed points are summarized in the  Table \ref{Table-fix} while the analyses are separately discussed for each fixed points below.

\begin{table}[h!]
\centering
\begin{tabular}{|c|c|c|c|c|c|c|}
  \hline
  Name & $x$ & $y$ & $z$  &$w_{eff}$ & existence & stability \\\hline
  (a) & 0 & 0 & $ 1$ & -1 &  $\gamma = 0$ & $0 \leq  \frac{s}{r} \leq  1 $\\\hline
  (b) & $\frac{1}{r}$ & 0 &  $1-\frac{1}{r}$ &  $ -1+\frac{s}{r}$ & $\gamma = \lambda$ & $\frac{\lambda}{1-\lambda} \leq  \frac{s}{r} < 0 $ \\\hline
  (c) & 0 & $1+ \frac{ s}{\lambda\,r}$ &  $-\frac{ s}{\lambda\,r}$ & $\frac{ s}{\lambda\,r}$ &   $\gamma = 1+ \frac{\lambda\,r}{s}$ & $\frac{\lambda}{1-\lambda} < \frac{s}{r} <-1 $  \\\hline
  (d) & $\frac{1}{1+\lambda(r-1)}$ & $\frac{\lambda(r-1)}{1+\lambda(r-1)}$  &  0 & $\frac{1}{\lambda -1}$ &  $\lambda = \frac{s}{s -r}$ &  $ 0<\lambda<1$  \\\hline
  (e) & $\frac{1+(\lambda-1)z_0}{1+\lambda(r-1)}$ & $-\frac{\lambda(1-r(z_0+1))}{1+\lambda(r-1)}$ &  $z_0$ & $\frac{1}{\lambda -1}$ & $\lambda= \gamma= \frac{s}{s -r}$ & $ 0<\lambda<1$ \\\hline
\end{tabular}
\caption{Summary of the properties of the fixed points.}\label{Table-fix}
\end{table}

\subsection{Fixed point (a)}
From Eq. (\ref{Eq-x}) and Eq. (\ref{Eq-y}),
it is obvious that the system has fixed point $(x, y) = (0,0)$. By using the constraint equations, one obtains  $z = 1$ and  $\gamma = 0$. This means that the function $P$ is constant and then this point corresponds to general relativity with a cosmological constant where the universe is dominated by the cosmological constant. To ensure such claim one can compute the corresponding effective equation of state parameter, which yields $w_{eff} = -1$.
This is exactly the equation of state parameter of the cosmological constant which drives the accelerating de-Sitter expansion.

The stability of the fixed point can be found by analyzing the eigenvalues of the linearly perturbed autonomous equations. By performing the linear perturbations, the eigenvalues can be written as $(\mu_1, \mu_2) = (-3 s/r, -3 -3 s/r)$. The stability requires both of the eigenvalues to be negative, or otherwise the fixed point is said to be unstable or saddle fixed points. In this case, the sign of those eigenvalues are determined by the value of the term $\frac{s}{r} = \left(\bar{X}-\eta\right)\frac{F_{,\bar{X}}}{G\eta}$ which means $0 \leq \frac{s}{r} \leq  1 $ for the stable fixed point. Note that in case of vanishing eigenvalues, like $s=0$, one has to consider the perturbations up to second order or use a numerical investigation in order to determine the stability. In this analysis, we ensure the stability in this case by the numerical method and we have found that it is stable.

Even though this fixed point can provide a period of late-time expansion, it is not much of interest due to the disappearance of the graviton mass. This resulting property is one of drawbacks
in the previous model of mass-varying massive gravity \cite{Huang:2012pe,Wu:2013ii,Leon:2013qh,Huang:2013mha}.

\subsection{Fixed point (b)}
One of possible fixed points may be in the form $(x, y) = (x_0,0)$ by which the universe is governed mainly by massive gravity alone.  From Eq. (\ref{Eq-x}), one can find $x_0$ as follows
\begin{eqnarray}
x_0 = \frac{1}{r}.
\end{eqnarray}
According to Eq. (\ref{Def-y}), there are two possible solutions for this kind of fixed point. One is $r=1$ in which $x_0=1,z_0=0$ and another one is $\lambda = \gamma$ in which $x_0 = \frac{1}{r}, z_0 = 1-\frac{1}{r}$.
The effective equation of state parameter can be written as
\begin{eqnarray}
w_{eff}=-1+\frac{F_{,\bar{X}} (\bar{X}-\eta) }{3G\eta}= -1 +\frac{s}{r}.
\end{eqnarray}
Interestingly, $w_{eff}=-1$ as $F_{,\bar{X}} = 0$ or $(\bar{X}-\eta) =0$. This characteristics is a usual cosmological solution of the original massive gravity. In particular, this condition indicates that the graviton mass ceases to vary, according to the Eq. (\ref{eqn:stuckeom}). Moreover, since in this case $z=1-\frac{1}{r}$, the pressure of the k-essence field is non-zero for $r > 1$ which means the k-essence field is supposed to be a matter with non-zero pressure (not dust).

In order to find the stability condition for this fixed point, one can find the eigenvalues of the linearly perturbed autonomous equations which can be written as
\begin{eqnarray}
(\mu_1, \mu_2)  =  \left(3 \frac{s}{r}, -3 +3 \frac{(\lambda-1)s}{\lambda\, r}\right).
\end{eqnarray}
Again, both of the eigenvalues contain the term $s/r$ and then the fixed point will be stable if $\frac{\lambda}{1-\lambda} \leq  \frac{s}{r} < 0 $. Note that, for this fixed point, it is possible to provide  $w_{eff} <-1$ to satisfy the observation which indicates that the mean value of the equation of state parameter is slightly less than $-1$ \cite{Ade:2013zuv}.

\subsection{Fixed point (c)}
One can obtain a fixed point such that $(x, y) = (0,y_0)$. From Eq. (\ref{Eq-x}), one can find $y_0$ as follows
\begin{eqnarray}
y_0  = 1+ \frac{s}{\lambda\,r}.
\end{eqnarray}
By using the constraint equation in Eq. (\ref{Eq-con1}), one obtains $z_0 =- \frac{s}{\lambda\,r}$. From the constraint equation in Eq. (\ref{Eq-con2}), we have
\begin{eqnarray}
\gamma =-\frac{y}{z}=-1+\frac{1}{z} = 1+\frac{1}{w_m},
\end{eqnarray}
where $w_m$ is the equation of state parameter of the fluid contributed from $P(X) = P_0 X^{(1+w_m)/2w_m}$.
The effective equation of state parameter can be written as
\begin{eqnarray}
w_{eff}=-z = \frac{s}{\lambda\,r}.
\end{eqnarray}
Again, There exist two significant branches of the solution such that this fixed point is a matter-dominated point.
If $z =0$, this corresponds to $w_{eff} =0$ which leads to the universe at matter dominated period.

The eigenvalues of the autonomous system can be written as
\begin{eqnarray}
(\mu_1, \mu_2)= \left(3+ 3\frac{s}{\lambda\,r},3- 3 \frac{s (\lambda -1)}{\lambda\,r }\right).
\end{eqnarray}
If one requires this point to represent the matter dominated epoch, one must put the parameters so that this point is unstable. This means the universe should evolve through this point to end up in other stable points since we know the matter dominated epoch should exist in the universe's timeline but not nowadays. One can see that, for small negative value of $s/r$, the universe can evolve in the standard history at which fixed point (c) corresponds to matter dominated period with $w_{eff} \sim 0$ and fixed point (b) corresponds to the late time expansion of the universe due to the contribution from the graviton mass.  However, it is not possible to alleviate the coincidence problem since the contribution of non-relativistic matter vanishes at  late time.

\subsection{Fixed point (d)}
According to the Eq. (\ref{Eq-x}) and Eq. (\ref{Eq-y}), one may consider the fixed point corresponding to the non-zero $x$ and $y$. This point can be obtained by evaluating both (non-zero) $x$ and $y$ from Eq. (\ref{Eq-con1}), Eq. (\ref{Eq-con2}), and Eq. (\ref{Eq-x}) while a constraint on the parameters by which the non-zero $(x,y)$ exist can be obtained from the Eq. (\ref{Eq-x}) and Eq. (\ref{Eq-y}). After simple manipulation, we have
\begin{align}
x &= \frac{1}{1+\lambda\left(r-1\right)}, \,\,\,\,\,\, y = \frac{\lambda\left(r-1\right)}{1+\lambda\left(r-1\right)}, \,\,\,\,\,\,\text{and } z = 0,
\end{align}
where $\gamma$ is arbitrary and $\lambda$ is fixed to be $\lambda = \frac{s}{s-r}$.
The effective equation of state parameter can be written as
\begin{align}
w_{eff} = \frac{1}{\lambda-1}.
\end{align}

To determine the stability of this point, we find the eigenvalues of the system of equations. Interestingly, this point renders both the autonomous equations to be degenerate. This can be seen by computing the linear perturbed equations for both $x$ and $y$ evaluated at this fixed point. The eigenvalues of this autonomous system are expressed as
\begin{align}
(\mu_1, \mu_2) = \left(0,\frac{3\lambda}{\lambda-1}\right). \label{eigend}
\end{align}
The vanishing eigenvalue here are nothing but an artifact of the degeneracy due to this fixed point. In particular, it is possible to redefine the variables such that the problem is reduced into one-dimensional system. With such redefinition, the stability of this fixed point is due to the non-zero eigenvalue in Eq. (\ref{eigend}), which can be negative when $0<\lambda<1$. If this condition is taken into account, requiring the fixed point (c) to represent the matter dominated era will restrict the combination $\frac{s}{r}$ to vanish.

This fixed point seems to provide a possible way to alleviate the coincidence problem due to the non-zero  $y$. However, it cannot be used since, at the late-time expansion, $w_{eff}$ must approach $-1$ and then leading to the fact that $(x, y) \rightarrow (1,0)$. Nevertheless, it  still provide an interesting result. For the case of $s = 0$ and $0<\lambda \ll 1$, this fixed point is stable while fixed point (b) is unstable and then we can use this fixed point as the one for the late-time expansion of the universe. For this condition the fixed point (c) is still  used for matter dominated period with $z = 0$. Therefore, this means that it is possible to obtain $z= 0$ for all history of the universe. This leads to the fact that, without  providing an extra non-relativistic matter field such as dark matter, the contribution from the graviton mass can play the role of both dark matter and dark energy. This is one of the crucial properties of this model since it can unify two main unknown contents of the universe; dark matter and dark energy by using only a graviton mass.

\subsection{Fixed point (e)}
Similarly to the derivation in fixed point (d), one can solve algebraic equation by imposing $\gamma = \lambda$ and requiring non-zero  $x$ and $y$.  As the result, the fixed point can be expressed as
\begin{align}
x &= \frac{1+(\lambda-1)z_0}{1+\lambda(r-1)}, \,\,\,\,\,\,  y = -\frac{\lambda(1-r(z_0+1))}{1+\lambda(r-1)},\,\,\,\,\,\, z = z_0,
\end{align}
where $\gamma = \lambda = \frac{s}{s-r}$ and $z_0$ is arbitrary. The effective equation of state parameter is the same as one in fixed point (d) which can be written as
\begin{align}
w_{eff} = \frac{1}{\lambda-1}.
\end{align}
Moreover, the eigenvalues for the stability analysis is still the same with the fixed point (d) and then the stability condition for this fixed point can be expressed as  $0 < \lambda < 1$. Even though this fixed point shares most properties with fixed point (d), it cannot provide the unification of two dark components
since $z$ must have a non-zero value.

From the above analyses, we experienced the incompatibility between a matter domination and a present dark energy domination. One may see that for a large $\lambda$, the fixed point (c) can represent the matter dominated epoch while the small value of $\lambda$ is needed in the fixed point (d) or (e) to solve the coincidence problem. It is natural to generalize the theory further by allowing $\lambda$ to change appropriately in time. This idea will be adopted and carefully analyzed in the next section.

\section{Extended analyses}\label{sec:exdyna}
As we have mentioned, even though the model can be used to unify the dark contents of the universe, it still cannot be used to solve the coincidence problem. According to our analysis, this is due to the fact that $\lambda$ is set to be a constant. In this section, we will show the possibility to solve the coincidence problem when $\lambda$ is set as a dynamical variable. For completeness, we will add  radiation into our consideration and then use numerical method to show that the radiation does not affect  the unification in the dark sector. Note that the equation of motion for the radiation is obtained by using the conservation of its energy momentum tensor or the continuity equation. By including the radiation and taking $\lambda$ as a dynamical variable, the autonomous equations can be written as
\begin{align}
x' &= 3 x \left(y+s x -\frac{s}{r} + \frac{4}{3} \Omega_r \right),\label{Eq-x2}\\
y' &= 3 y\left(y+s x -1-\frac{s}{\lambda r} + \frac{4}{3} \Omega_r \right),\label{Eq-y2}\\
\Omega_r' &= 3 \Omega_r \left(y+s x + \frac{4}{3} (\Omega_r -1)\right),\label{Eq-Or}\\
\lambda' &= \frac{6 s}{r}\left(\frac{\lambda}{2} -(1+\Gamma)\right),\label{Eq-lambda2}\\
1 &= x+ y + z + \Omega_r, \label{Eq-con12}\\
y &=  -\lambda x (1-r) - z \gamma,\label{Eq-con22}\\
\Omega_r &\equiv \frac{\rho_r}{3M^2_p H^2},\label{Def-Or}
\end{align}
where $\rho_r$ is the energy density of the radiation. The effective equation of state  parameter can be written as
\begin{eqnarray}
w_{eff}  = -1 + y +x s +\frac{4}{3} \Omega_r.
\end{eqnarray}

\begin{figure}[h!]
\includegraphics[scale=0.45]{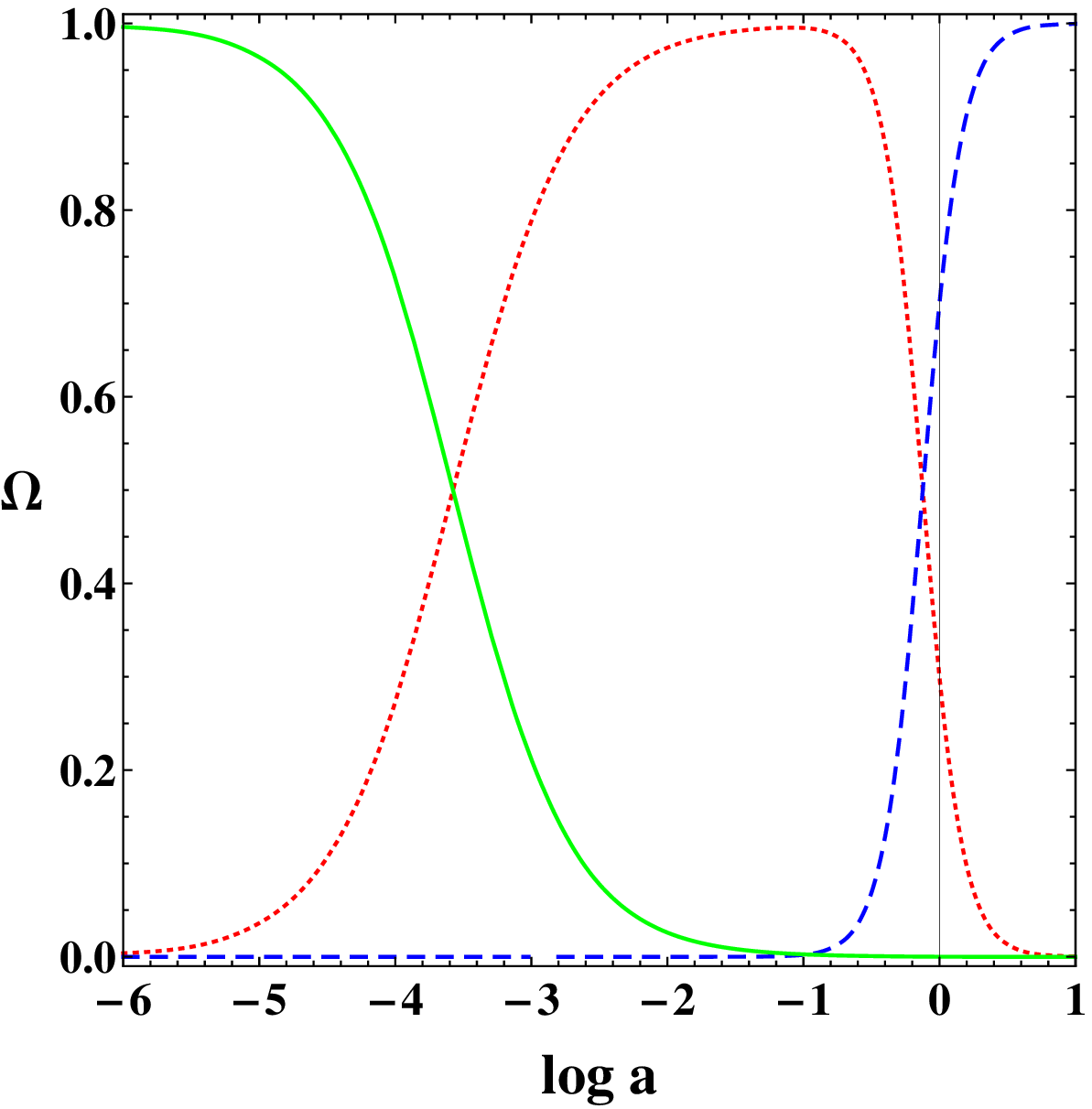}\qquad\qquad
\includegraphics[scale=0.45]{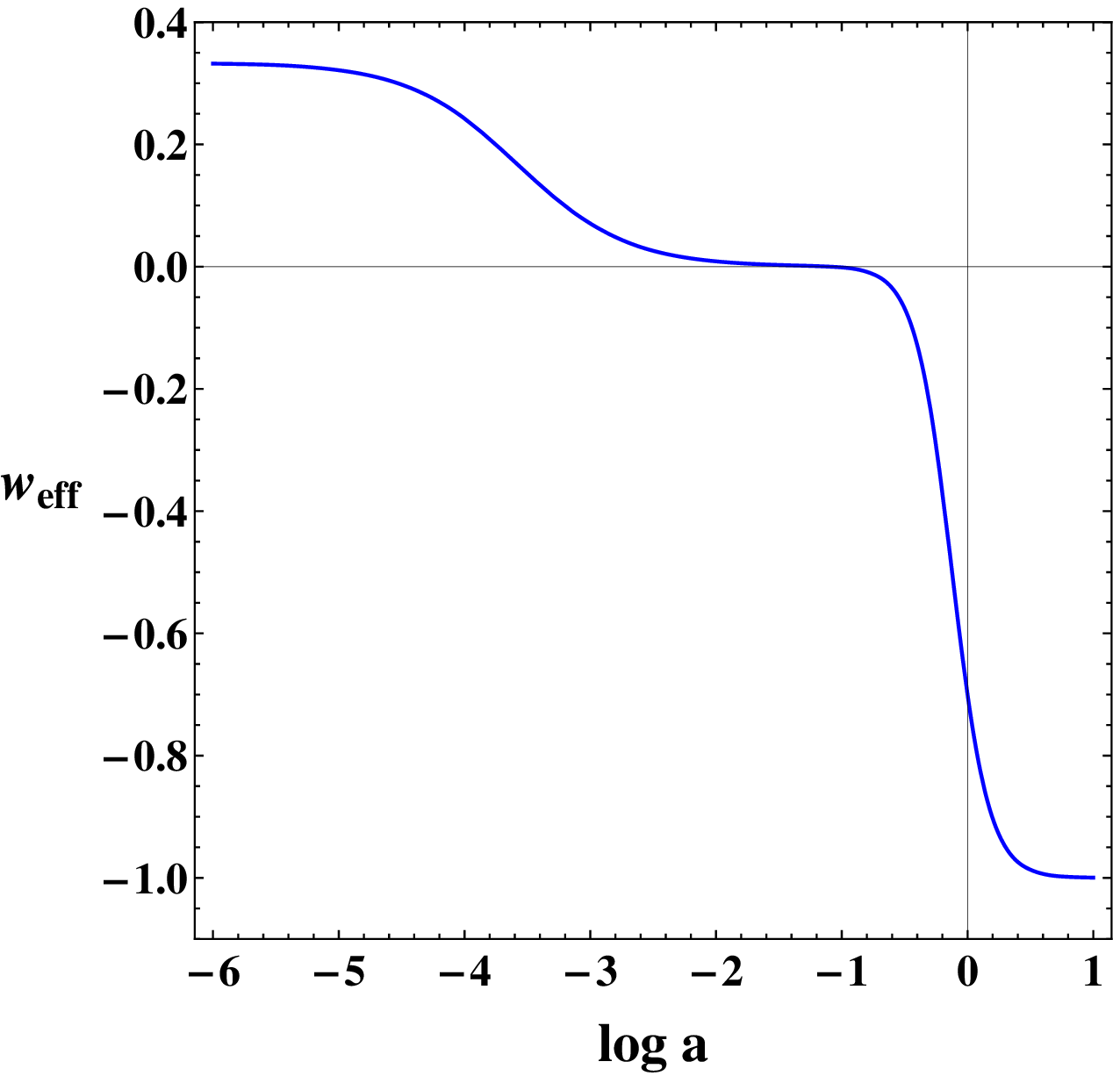}
{\caption{\label{no-z} The left panel shows the evolution of $x$, $y$ and $\Omega_r$. The dotted-red line  represents the evolution of $x$, the dashed-blue line represent the evolution of $x$ and the solid-green line represents the evolution of $\Omega_r$. For the right panel, the evolution of $w_{eff}$} is represented.}
\end{figure}

From Eq. (\ref{Eq-Or}), we can see that all fixed points we found in the previous section still exist with $\Omega_r = 0$. Also, there exists the unstable fixed point such that $\Omega_r = 1$ while $x$ and $z$ (hence $y$) vanish. From Eq. (\ref{Eq-lambda2}), one can see that $\lambda$ does not couple to the others and the fixed point takes place at $\lambda = 2(\Gamma +1)$. For simplicity, one can set $\Gamma$ as a constant. In order to confirm the claim in the previous section such that there exists the standard evolution without introducing k-essence Lagrangian or in the case of  $z=0$, we use numerical method to evaluate the equations above by  setting $s = 0$. The evolutions of $x$, $y$ and $\Omega_r$ are illustrated in the left panel of Fig. \ref{no-z} and the evolution of the effective equation of state parameter shown in the right panel of Fig. \ref{no-z}. We can see that there exists the non-relativistic matter inferred as dark matter represented by the variable $y$ while the variable $x$ represents the dark energy that drives the late-time expansion of the universe. Both $x$ and $y$ are contributed from the graviton mass.

\begin{figure}[h!]
\includegraphics[scale=0.46]{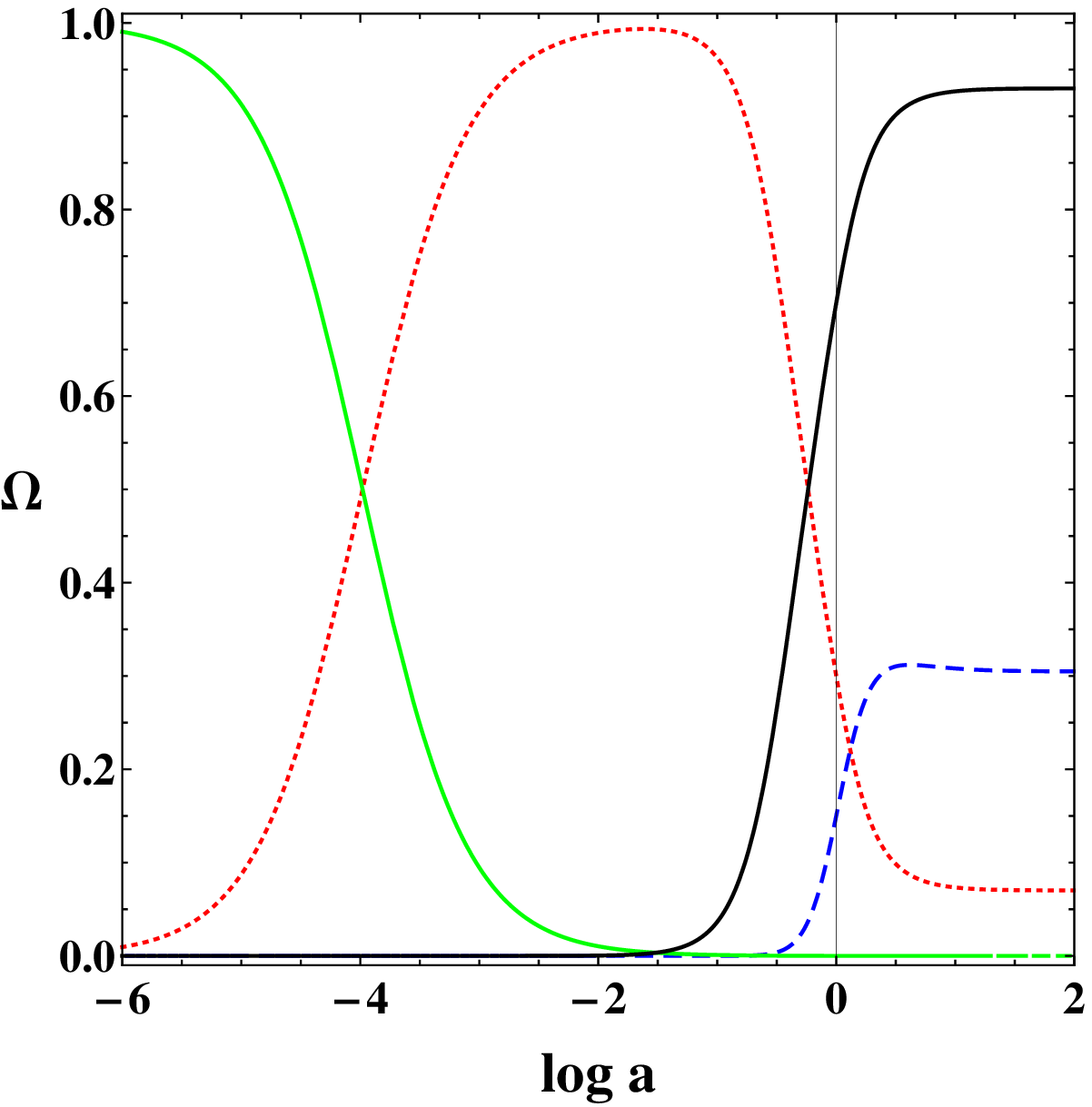}\qquad\qquad
\includegraphics[scale=0.45]{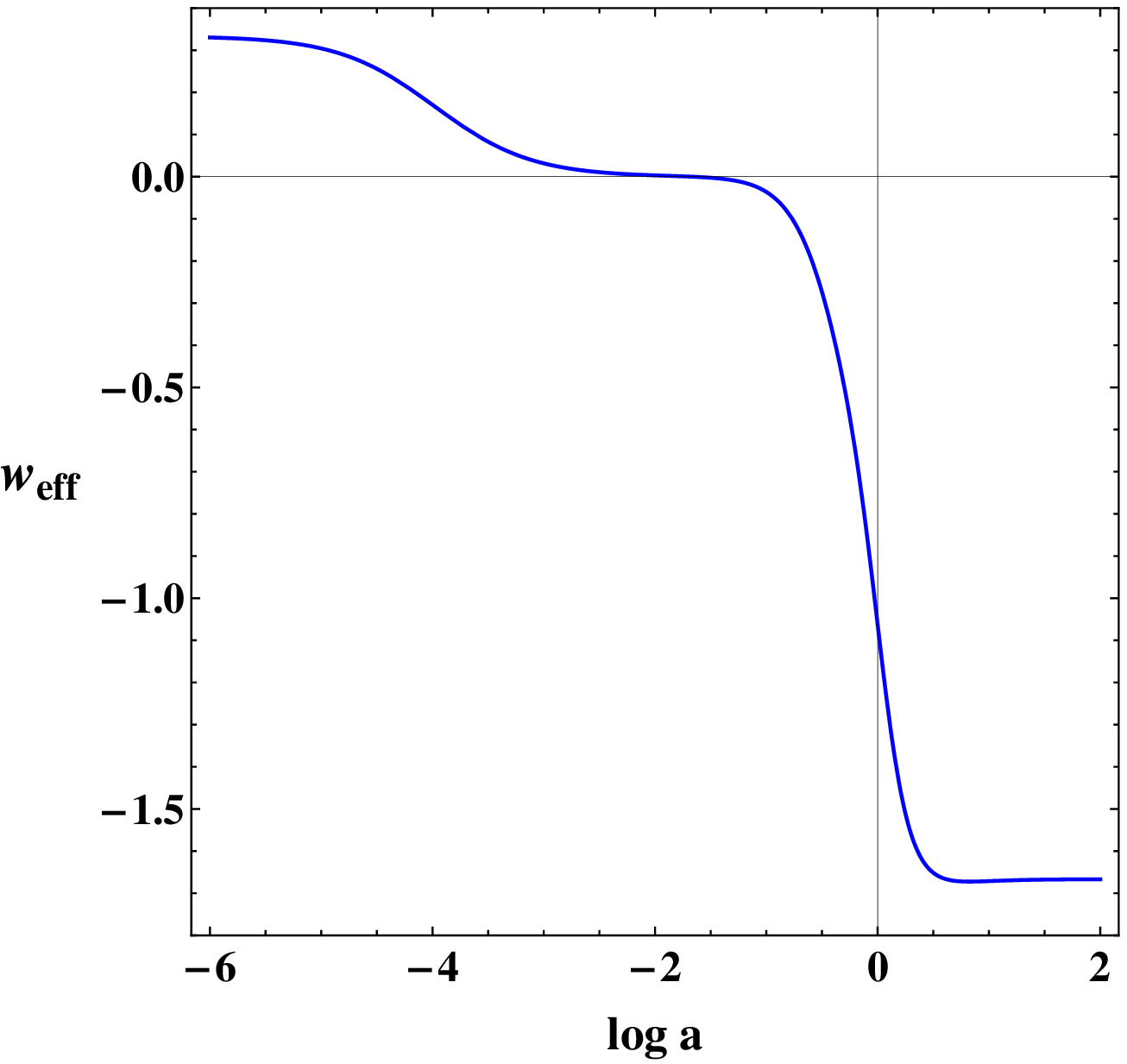}
{\caption{\label{lng} The left panel shows the evolution of $x$, $y$, $x+z$ and $\Omega_r$. The dotted-red line  represents the evolution of $x$, the dashed-blue line represent the evolution of $x$, the solid-black line represents the evolution of $x+z$ and the solid-green line represents the evolution of $\Omega_r$. For the right panel, the evolution of $w_{eff}$} is represented. We set the parameters such that $\lambda_f = 0.4$ and $\lambda_0 = 1.0$ where $\lambda_f$ is the value at the fixed point and $\lambda_0$ is one at the present time.}
\end{figure}

Now, let us consider the possibility to solve the coincidence problem. Let us use the fixed point (e) to be one corresponding to the late-time expansion of the universe. For this fixed point, the parameters $s, r$ and $\Gamma$ are obtained by giving the initial conditions for the dynamical variables. In order to obtain the dynamics of all variables, we have to put the initial conditions slightly away from the fixed point. It is sufficient to put $\lambda$  slightly above the fixed point since we need $\lambda$ to grow as time goes backward to ensure that it have enough value for matter-dominated period. In order to obtain $w_{eff} \sim -1$ at the present time, we have to set the value of the variable $\lambda$ at the fixed point as $\lambda_f \rightarrow 0$. As a result, $\frac{s}{r} = \frac{\lambda_f}{\lambda_f - 1} \rightarrow 0$. In order to obtain a proper matter-dominated period, one has to put the initial value of $\lambda$ far away from the fixed point.
This situation makes the fixed point (b) stable  and then the
system evolves to the point (b) eventually. Therefore, in order to  have the fixed point (e) at late time, one has to set $w_{eff}$ below $-1$ at the fixed point so that the point (e) becomes a stable point. According to this setting, we show the evolution of the dynamical variables reaching the fixed point (e) to alleviate the coincidence problem in Fig \ref{lng}. Note that we set $\lambda_f =0.4$ leading to $w_{eff} = -1.67$  and $\lambda_0 =1.0$.

\begin{figure}[h!]
\includegraphics[scale=0.45]{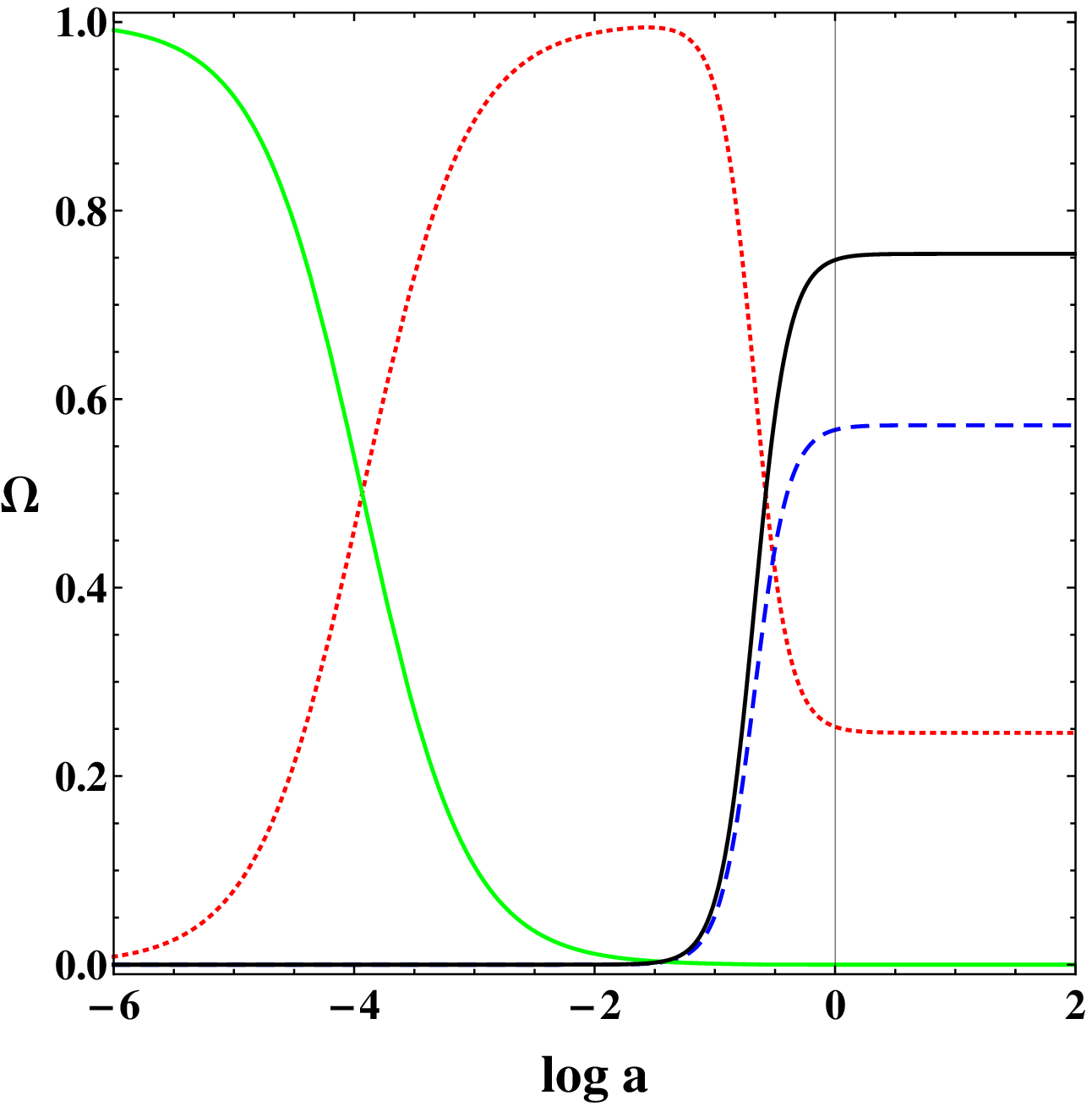}\qquad\qquad
\includegraphics[scale=0.47]{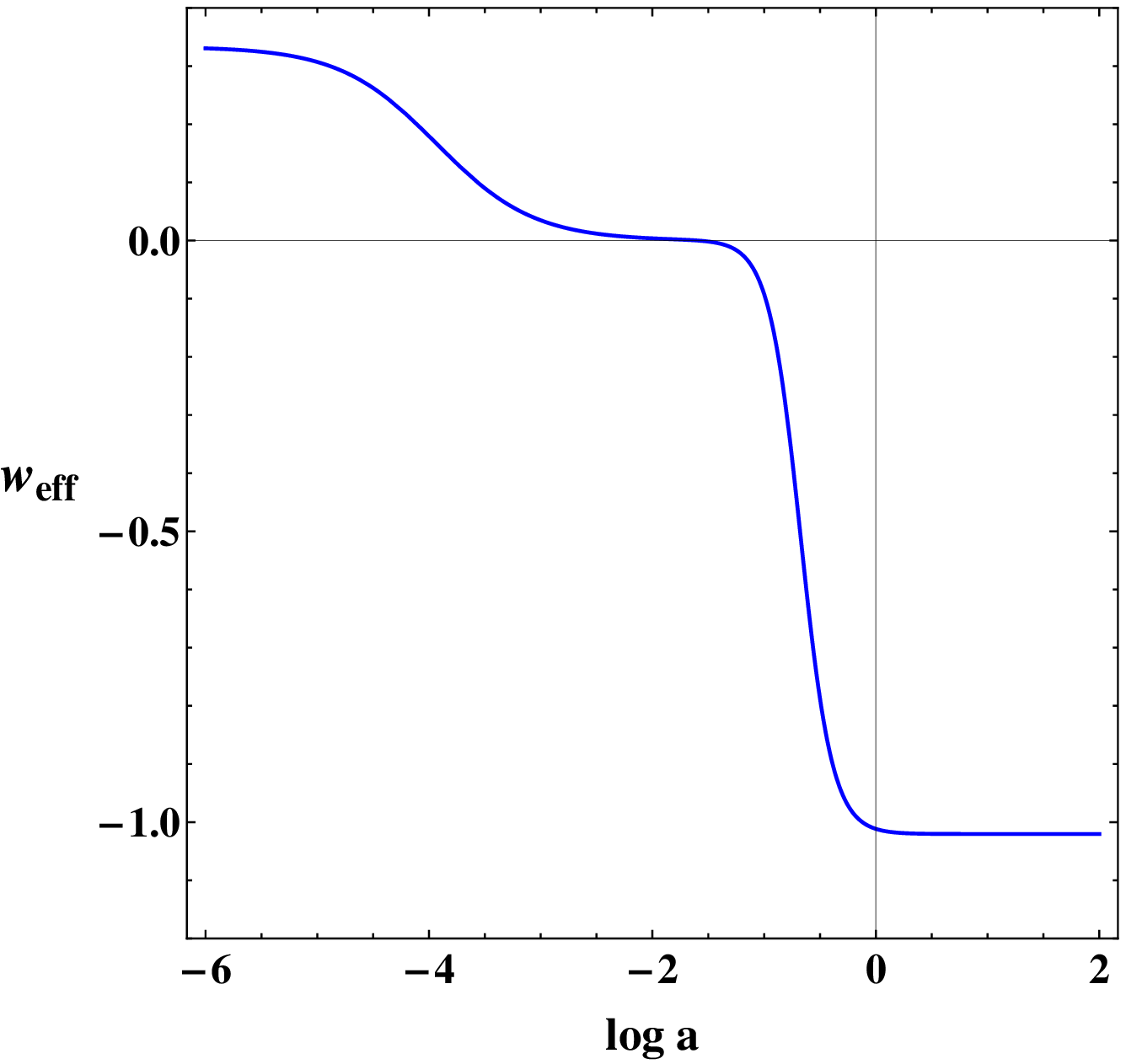}
{\caption{\label{leg} The left panel shows the evolution of $x$, $y$, $x+z$ and $\Omega_r$. The dotted-red line  represents the evolution of $x$, the dashed-blue line represent the evolution of $x$, the solid-black line represents the evolution of $x+z$ and the solid-green line represents the evolution of $\Omega_r$. For the right panel, the evolution of $w_{eff}$} is presented.}
\end{figure}

In order to overcome the incompatibility among the fixed points, one may extend the analysis by allowing $s$, $\Gamma$ or $r$ to be dynamical variables. This will make the dynamical system more complicated. We found another possibility to overcome this incompatibility by imposing the constraint $\lambda = \gamma$ for the entire evolution.
As a result, we have only three independent equations for six variables and three constraints. The dynamical variable $\lambda$ can be written in terms of other variables as
\begin{align}
\lambda &= \frac{y}{r x +y +\Omega_r -1}.
\end{align}
As a result, by setting the initial condition at the radiation dominated period, the evolution of the dynamical variables and the effective equation of state are shown in Fig. \ref{leg}. From this figure, one can see that the evolution of the universe reaches the fixed point (e) at  late time while the matter and radiation period are also properly presented. For the plot in this figure, we set $\lambda_f = 0.02$ and then the consequent results are $\Gamma = -0.99$ and $w_{eff} \sim -1.02$. Note that the behavior of resulting plot in Fig \ref{leg}. is sensitive to the initial value of $x$ at the radiation dominated period where we set it as $x_i \sim 10^{-16}$.

\section{Conclusion}

We have constructed a new class of the mass-varying massive gravity in which not only the k-essence field but also its kinetic term  determine the variation of the graviton mass. We have shown in section \ref{sec:eom} that there is a possibility for the graviton mass to live at late time compared with the previous model whose the graviton mass only depends on the scalar field and shrinks as the universe grows \cite{Huang:2012pe,Wu:2013ii,Leon:2013qh,Huang:2013mha}. After simple manipulations and under particular assumptions, we found that a ``dust-like'' matter which behave like a non-relativistic dust can naturally comes out from the graviton mass and it is a possible candidate for a dark matter. This can be seen more clearly in the case $P=0$ in which the dark matter comes solely from the varying graviton mass. Having such matter in the system, this model of massive gravity can describe the cosmic accelerating expansion with the equation of state parameter close to $-1$ while the universe is not entirely dominated by the dark energy part contributed also by the graviton mass. This property signals a possibility of having the universe composed of comparable amounts between dark energy and dark matter, known as the cosmic coincidence problem. To obtain a finer description on this, the usual method of the dynamical analysis is performed by taking the dark matter candidate into account and the results are carefully investigated on the issue of the coincidence problem. For the first simple case, the exponent of the kinetic term in the graviton mass $\lambda$ is kept to be constant. We found the fixed points which correspond to various epochs in the history of the universe such as the matter-dominated period and massive-gravity-dominated periods. However, to have those fixed points with the appropriate stabilities in the evolution of the universe, the results suggest a system with $\lambda$ as additional variable. The more general case, where $\lambda$ is allowed to vary, is investigated where the radiation is included. While the result covers all the fixed points in the constant $\lambda$ case, this allows the evolution in which there exist a matter-dominated period as well as a late-time expansion epoch. There are several crucial points in this investigation. Firstly, we obtain the universe in which the graviton mass serves as both dark energy  and  dark matter while it can still drive the cosmic acceleration. Secondly, to solve the coincidence problem, we obtain a universe with the effective equation of state parameter significantly below $-1$ unless both $\lambda$ and $\gamma$ are set to be equal with one another for the entire evolution of the universe. Since the analyses are under particular assumptions, this model still has rooms to be studied in a more complicated way. For example, one can exclude the assumptions proposed in this work for a more complex system or one can consider this model in a different aspect like its astrophysical implications. Not only on the applications, but studying on the theoretical consistency, whether there exists a ghost instability or not, is also a worthy challenge which we leave it as a future work. Apart from those mentioned, one may think of constraining the model with various observations. This idea is also interesting since the observations may judge the fate of this model by tightening it with constraints.

\begin{acknowledgments}
P.W is supported by Thailand Research Fund (TRF) through grant TRG5780046. L.T. is supported by the Faculty of Science, Mahidol University through Sritrang-Thong Ph.D. scholarship. Moreover, the authors would like to thank String Theory and Supergravity Group, Department of Physics, Faculty of Science, Chulalongkorn University for hospitality during this work was in progress.
\end{acknowledgments}

\appendix

\end{document}